\documentclass[onecolumn]{jpsj3}%
\usepackage{amsfonts}
\usepackage{amsmath}
\usepackage{amssymb}
\usepackage{graphicx}%
\newcommand{\PR}[1]{Phys. Rev. B {\bf {#1}}}

\newcommand{\PRL}[1]{Phys.\ Rev.\ Lett. {\bf {#1}}}
\newcommand{\JPSJ}[1]{J.\ Phys.\ Soc.\ Jpn. {\bf #1}}
\newcommand{\sgt}{$\raisebox{-0.6ex}{$\stackrel{>}{\sim}$}$}
\newcommand{\slt}{$\raisebox{-0.6ex}{$\stackrel{<}{\sim}$}$}
\newcommand{\bvec}[1]{\mbox{\boldmath $#1$}}
\newcommand{\D}{\delta }

\newcommand{\vk}{\bvec{k}}
\newcommand{\vS}{\bvec{S}}
\newcommand{\vr}{\bvec{r}}
\newcommand{\eq}[1]{eq.~(\ref{#1})}
\newcommand{\fig}[1]{Fig.~\ref{#1}}

\newcommand{\be}{\begin{equation}}
\newcommand{\ee}{\end{equation}}
\newcommand{\bea}{\begin{eqnarray}}

\newcommand{\eea}{\end{eqnarray}}
\newcommand{\bean}{\begin{eqnarray*}}

\newcommand{\bfi}{\begin{figure}}
\newcommand{\efi}{\end{figure}}
\newcommand{\bc}{\begin{center}}
\newcommand{\ec}{\end{center}}

\newcommand{\lsim}{ < \kern -11.8pt \lower 5pt \hbox{$\displaystyle \sim$}}
\newcommand{\gsim}{ > \kern -12pt   \lower 5pt   \hbox{$\displaystyle \sim$}}

\begin{document}

\title{Instability toward Formation of  Quasi-One-Dimensional Fermi Surface \\ 
in Two-Dimensional $t$-$J$ Model}
\author{Hiroyuki Yamase$^{1}$ and Hiroshi Kohno$^{2}$}

\inst{$^{1}$Institute for Solid State Physics, University of Tokyo, 
 5-1-5 Kashiwanoha, Kashiwa, Chiba 277-8581 \\ 
$^{2}$Graduate School of Engineering Science, 
Osaka University, Toyonaka, Osaka 560-8531}

\recdate{February 16, 2000}

\abst{We show within the slave-boson mean field approximation that  
the two-dimensional  $t$-$J$ model has an intrinsic instability toward 
forming a quasi-one-dimensional (q-1d) 
Fermi surface.  This q-1d state 
competes with,  and is overcome by, the $d$-wave pairing state 
for a realistic parameter choice. 
However, we find that a small spatial anisotropy in $t$ and $J$ exposes 
the q-1d instability which has been  
hidden behind the $d$-wave pairing state, and brings 
about  
the coexistence with the $d$-wave pairing. 
We argue that this coexistence can be realized 
in La$_{2-x}$Sr$_{x}$CuO$_4$ systems.}

\kword
{two-dimensional $t$-$J$ model, mean field approximation, 
Fermi surface, $d$-wave pairing state, quasi-one-dimensional state,  
competition,  coexistence,  orthorhombicity, LSCO}

\maketitle

\section{Introduction}
Elastic neutron scatterings in 
La$_{1.6-x}$Nd$_{0.4}$Sr$_{x}$CuO$_4$ 
(LNSCO)\cite{tranquada1,tranquada2,tranquada3} have revealed 
that static charge density 
modulation (CDM) coexists with static 
incommensurate antiferromagnetic long-range order even in the
superconducting state. 
This coexistence has often been  
discussed in terms of the so-called 
\lq spin-charge stripe model'\cite{tranquada1,tranquada2}. 
Direct experimental evidence confirming this model, however, has not
been obtained so far. 
On the theoretical side, it is still controversial 
on a point whether the $t$-$J$ model has the \lq spin-charge stripe'
ground state\cite{white1,white2,white3,hellberg1,hellberg2}.

On the other hand,  
we proposed\cite{yamase} a quasi-one-dimensional (q-1d) picture of the 
Fermi surface (FS) in La$_{2-x}$Sr$_{x}$CuO$_4$ (LSCO). 
It was motivated by the apparently contradicting experimental results between 
the angle-resolved photoemission spectroscopy (ARPES)\cite{ino} 
and the inelastic neutron scattering\cite{yamada} on one hand, 
and by our theoretical finding\cite{yamase3} that the two-dimensional (2d) 
$t$-$J$ model has 
an intrinsic instability toward forming a q-1d FS 
on the other hand.  

In this paper, 
we report a detailed analysis of the latter, namely on the 
intrinsic instability of the 2d $t$-$J$ model toward forming a q-1d FS,  
which can be regarded as a microscopic support of the proposed q-1d
picture\cite{yamase}. 
For a realistic parameter choice, however, 
this q-1d state proves to compete with, and be overcome by,  
the $d$-wave pairing 
state (the $d$-wave singlet resonating-valence-bond ($d$-RVB) state).  
Nonetheless, a small spatial anisotropy in $t$ and $J$ exposes 
the q-1d instability which has been hidden behind the $d$-RVB,  and  
brings about the coexistence with the $d$-RVB state.  
We argue that this coexistence can be realized in LSCO systems. 
We note that charge distribution is homogeneous in the present q-1d state
and that any relation to the \lq spin-charge stripe model' 
has not been obtained at present. In the following, we describe the model 
and the calculation scheme in \S 2, and results in \S 3. 
Discussions are given in  \S 4.

\section{Model}
As a theoretical model for high-$T_{\rm c}$ cuprates, 
we use the 2d ({\it spatial  isotropic}) $t$-$J$ model defined 
on a square lattice: 
\bea
 & &H = -  \sum_{i,\,j,\, \sigma} t\,^{(l)}_{i\, j}
 f_{i\,\sigma}^{\dagger}b_{i}b_{j}^{\dagger}f_{j\,\sigma} + 
   \sum_{<i,j>} J_{i j} \vS_{i} \cdot \vS_{j}, \label{t-J} \\
& &\hspace{2mm} \sum_{\sigma}f_{i\,\sigma}^{\dagger}f_{i\,\sigma}
          +b_{i}^{\dagger}b_{i}=1  
  {\rm \quad  at\  each\  site}\ i, \label{constraint}
\eea  
where $f_{i\,\sigma}$ ($b_{i}$) is a fermion (a boson) operator  
that carries spin $\sigma$ (charge $e$), namely the 
so-called slave-boson scheme, and  
$t^{(l)}_{i\, j}=t^{(l)}$ is a hopping integral between 
the $l$-th nearest neighbor (n.n.) sites $i$ and $j$ ($l\leq 3$), 
$J_{i j}=J >0$ is the superexchange coupling between the n.n. spins, and 
$\vS_{i} = \frac{1}{2}\sum_{\alpha,\beta}
   f_{i\,\alpha}^{\dagger}\bvec{\sigma}_{\alpha\,\beta}f_{i\,\beta}$
 with Pauli matrix $\bvec{\sigma}$.  
The constraint \eq{constraint} excludes double occupations.  
(Later, we will consider the {\it anisotropic} $t$-$J$ model in the
 sense that a spatial anisotropy is introduced in $t^{(1)}_{i\, j}$ and
 $J_{i j}$ in \eq{t-J}. See eqs. (\ref{LTT1}) and (\ref{LTT2}).) 

Following the previous procedure\cite{tanamoto}, we introduce mean
fields: 
$\chi_{\tau}^{(l)}
\equiv \left< \sum_{\sigma}f_{i\,\sigma}^{\dagger}
f_{i+\tau \,\sigma}\right>$, 
$\left<b_{i}^{\dagger}b_{i+\tau}\right>$ and 
$\Delta_{\tau}^{(1)}  
\equiv \left<f_{i\,\uparrow}f_{i+\tau \,\downarrow}- 
f_{i\,\downarrow}f_{i+\tau \,\uparrow}\right>$, 
where each is taken to be a real constant independent of lattice 
coordinates, but is allowed its dependence 
on the bond direction $\bvec{\tau}=\vr_{j}-\vr_{i}$ (see \fig{Xij}). 
Also, the local constraint \eq{constraint} is loosened to a global one,   
$\sum_{i}\left(\sum_{\sigma}f_{i\,\sigma}^{\dagger}f_{i\,\sigma}
          +b_{i}^{\dagger}b_{i}\right)=N$ with $N$ being the total 
number of lattice sites. 
We then decouple the Hamiltonian \eq{t-J} to obtain 
\bea
&&H_{\rm MF} = 
\sum_{\vk, \,\sigma}\xi_{\vk} f_{\vk\,\sigma}^{\dagger} f_{\vk\,\sigma}
+ \sum_{\vk}\left(\Delta_{\vk} f_{-\vk\,\downarrow}^{\dagger} 
f_{\vk\,\uparrow}^{\dagger} + 
\Delta_{\vk}^{\ast} f_{\vk\,\uparrow}
f_{-\vk\,\downarrow} \right), \label{HMF}\\
& &\xi_{\vk}=-2\sum_{l,\,\tau}F_{\tau}^{(l)}\cos k_{\tau} -\mu \; ,\\
& &F_{\tau}^{(l)}=t\!\,^{(l)}
\left<b_{i}^{\dagger}b_{i+\tau}\right> 
        +\frac{3}{8}J \chi_{\tau}^{(l)} 
 \; \D_{l,\,1} \; ,   \label{F} \\
& &\Delta_{\vk}=-\frac{3}{4}J 
\left( \Delta_{x}^{(1)} \cos k_{x}
+\Delta_{y}^{(1)} \cos k_{y} \right),
\label{singlet}
\eea 
where $\mu$ is the chemical potential, $\D_{l,1}$ is the 
Kronecker's delta, and 
$k_{\tau}=k_{x}$ or $k_{y}$ for $l=1$, $k_{\tau}=k_{x}+k_{y}$ or 
$k_{x}-k_{y}$ for $l=2$ and  $k_{\tau}=2k_{x}$ or $2k_{y}$ for $l=3$. 
We approximate bosons to be Bose-condensed and 
neglect the kinetic term for bosons in \eq{HMF}. 
This approximation will be reasonable at low temperature, 
and leads to $\left<b_{i}^{\dagger}b_{i+\tau}\right> \approx \D$  
where $\D$ is the hole density. 
It is to be noted that we do not assume four-fold symmetry, 
$\chi_{x}^{(1)} =
 \chi_{y}^{(1)}$ and  
$\left| \Delta_{x}^{(1)}\right| =
\left| \Delta_{y}^{(1)}\right|$, which 
was assumed previously\cite{tanamoto}.  
In the following, we abbreviate $\chi_{\tau}^{(1)}$ 
and $\Delta_{\tau}^{(1)}$ to $\chi_{\tau}$  and $\Delta_{\tau}$, respectively.

\section{Results}
In \S 3.1 and \S 3.2,  focusing our attention on the LSCO systems, we set 
the parameters as $t^{(1)}/J =4$, $t^{(2)}/t^{(1)}=-1/6$ and 
$t^{(3)}/t^{(1)}=0$, and  determine the mean fields by minimizing the
free energy. 
These parameters reproduce the
observed FS at $\D=0.30$\cite{ino} in LSCO\cite{xD}.  
We also study with the other parameter choice in \S 3.3. 
\subsection{Isotropic $t$-$J$ model}
\subsubsection{Numerical calculations}
We first show the numerical results obtained under the constraint 
$\Delta_{\tau}=0$. 
Figure \ref{Xq1d}(a) shows $\chi_{\tau}$ 
as a function of temperature $T$. A second-order 
phase transition takes place at $T=T_{\rm q1d}$, below which the 
four-fold symmetry of 
$\chi_{\tau}$ is broken 
spontaneously, that is 
$\chi_{x} \ne
\chi_{y}$.  
The 2d FS (gray line in \fig{Xq1d}(b)) at high temperature changes into the 
q-1d FS (solid line) for  $T < T_{\rm q1d}$. 
Figure \ref{Tq1d} shows $T_{\rm q1d}$ as a function of $\D$. 
The q-1d state is realized below the critical doping rate, 
$\D_{\rm q1d}\approx 0.13$. The jump of $T_{\rm q1d}$ at $\D_{\rm q1d}$ 
indicates a weak first-order phase transition at $T=0$ as a function of $\D$.

When we remove the constraint 
$\Delta_{\tau}=0$, 
the 2d $d$-RVB state ($\Delta_{x}=- \Delta_{y}$) 
sets in before       
the q-1d instability occurs, and the 
q-1d state does not appear.

\subsubsection{Ginzburg-Landau analysis}  
To see the origin of the q-1d state and its competition with 
the $d$-RVB, we examine 
a Ginzburg-Landau (GL) free energy.  
Under the constraint $\Delta_{\tau}=0$,  
we vary $\chi_{\tau}$ and $\mu$ infinitesimally  
around the isotropic 2d state, 
$\chi_{0}$ and $\mu_{0}$, 
keeping $\D$ fixed: $\chi_{x} = \chi_{0} + \D\! \chi$,  
$\chi_{y} = \chi_{0} - \D\! \chi$, and  $\mu = \mu_{0}+\D\!\mu$. 
Up to the second order in $\D\! \chi$ and $\D\!\mu$, we 
estimate the dominant terms in the GL free energy as 
\be
F-F_{0} \sim \frac{3 J}{4} (1-a) (\D\!\chi)^{2} \label{GL}. 
\ee
Here $F_{0}$ is the free energy in the  isotropic 2d state and  
\be
a=\frac{3 J}{4}\frac{1}{N}\sum_{\vk} 
\left( - \frac{\partial n_{F}} {\partial  \xi_{\vk}}\right) 
\left(\cos k_{x} -\cos k_{y}\right)^{2} >0 \; ,  \label{a}
\ee
where  $n_{F}$ is the Fermi-Dirac distribution function. 
The GL coefficient, $1-a$, at $\D=0.05 $ is 
shown in \fig{1-a} as a function of $T$.
It becomes negative below $T_{\rm q1d} \approx 0.09 J $, 
signaling an instability toward the q-1d state. 
This value of $T_{\rm q1d}$ is the same as that 
shown in \fig{Xq1d}(a), which confirms that the q-1d 
instability is controlled by $a$. 

Since in \eq{a}, 
the factor $- \frac{\partial n_{F}} {\partial  \xi_{\vk}}$ limits 
$\vk$ to a region close to the FS, 
and the form factor $(\cos k_{x} -\cos k_{y})^{2}$  
takes maxima at points $(\pi,\, 0)$ and 
$(0,\,\pi)$, the condensation energy for the q-1d state comes 
mainly from  fermions on the FS near $(\pi,\, 0)$ and $(0,\,\pi)$. 
The same energetics  holds  for the $d$-RVB state also.    
In this sense, the q-1d state competes with the $d$-RVB state. 
Figure \ref{energy gain} shows that the condensation energy is larger for 
the latter. 
This is why  the $d$-RVB state has overcome the q-1d state 
in our numerical calculation.

\subsection{Anisotropic $t$-$J$ model}
Having seen that the q-1d state has free energy higher than the $d$-RVB 
state, we next ask a question: is there any perturbation which 
favors the q-1d state relative to the $d$-RVB state and  
stabilizes the q-1d state, or at least the coexistence 
with the $d$-RVB state?  
We here show that 
a small spatial anisotropy in $t^{(1)}$ and $J$ 
exposes the q-1d instability which has  been hidden 
behind  the $d$-RVB,  and  brings  about the coexistence 
with the $d$-RVB state.  
As an origin of this 
anisotropy, we consider the low-temperature tetragonal (LTT) structure  
and introduce as\cite{bruce,yamase2} 
\bea
& &t_{x}^{(1)} = t^{(1)}, \quad 
t_{y}^{(1)} = t^{(1)} (1-3.78 \tan^{2}\theta), \label{LTT1}\\
& &J_{x} = J, \quad 
J_{y} = J (1-2 \cdot 3.78 \tan^{2}\theta), \label{LTT2}
\eea
where $\theta$ is a tilting angle of the CuO$_{6}$ octahedra and the 
subscripts, $x$ and $y$, indicate the bond direction. 
(In \S4.1.1, we will discuss a possible origin of this 
anisotropy in LSCO whose 
crystal structure is the low-temperature orthorhombic (LTO).)    
Taking 
$\theta=5^{\circ}$\cite{radaelli}, namely 
$t_{y}^{(1)}/t_{x}^{(1)} \approx 0.97$ and 
$J_{y}/J_{x} \approx 0.94$,   
we determine the mean fields 
without the constraint $\Delta_{\tau}=0$. 

Figure \ref{Xq1ds}(a) shows the degree of the anisotropy, 
$\frac{\chi_{x}-\chi_{y}}
{\chi_{x}+\chi_{y}}$,  
as a function of $T$.   
For $\D \slt 0.20$, the anisotropy 
is largely enhanced 
as decreasing $T$ and after showing 
a cusp at $T_{\rm RVB}$, onset temperature of the $d$-RVB, 
it decreases but approaches to a still enhanced  
value as $T \rightarrow 0$. 
The degree of the anisotropy at 
$T \sim 0.5 J$ hardly depends on $\D$ and hence can be 
solely due to the given 
anisotropy in $t^{(1)}$ and $J$. 
The enhanced anisotropy at lower temperature  
comes from the  intrinsic q-1d instability, 
whose competition with the $d$-RVB 
makes the cusp at $T_{\rm RVB}$.  
This competition also 
suppresses the value of $T_{\rm RVB}$ about $6 \sim 8 \%$ 
compared to that 
of $T_{\rm RVB}$  for the (pure) $d$-RVB state realized in the 
{\it isotropic} $t$-$J$ model.   
Despite the competition, a  still enhanced anisotropy survives at 
$T = 0$ and becomes smaller as increasing $\D$. 
Figure \ref{Xq1ds}(b) shows the FSs at $T=0.001J$ 
for $\D=0.05$ and $0.15$, which are q-1d.  
For $\D \sgt 0.25$, the value of 
$\frac{\chi_{x}-\chi_{y}}
{\chi_{x}+\chi_{y}}$ 
does not depend on $T$ appreciably. 
This  behavior qualitatively different from that for $\D \slt 0.20$ 
can be understood as coming from the fact 
that  the intrinsic q-1d instability is limited to  
$\D \slt \D_{\rm q1d} \approx 0.13$ in the {\it isotropic} $t$-$J$ model 
as  found under 
the constraint $\Delta_{\tau}=0$.  
In this sense, the value of $\D_{\rm q1d}$ is 
a rough measure of the extent 
of $\D$ where the intrinsic q-1d instability appears in the 
{\it anisotropic} $t$-$J$ model.

We note that in the coexistent state an extended $s$-wave component, 
$\Delta_{s}$, mixes into  the $d$-wave component, $\Delta_{d}$: 
\bea
& & \Delta_{d}= \frac{1}{2}\left|
\Delta_{x}-\Delta_{y}\right|, \\ 
& & \Delta_{s}= \frac{1}{2}\left|
\Delta_{x}+\Delta_{y}\right|.
\eea
Figure \ref{s-wave} shows that the mixing is  
about 1.5\% for $\D \slt 0.15$. This small $s$-wave ratio 
does not  shift the Fermi point ($d$-wave node) appreciably from 
the symmetry axis $k_{y}=\pm k_{x}$; 
its shift is  less than $\sim 0.1\%$ of the 1st Brillouin zone.   

\subsection{Band parameter dependence}
Next we examine the band parameter dependence of the q-1d instability. 
Taking $t^{(1)}/J=4$ in common, we  
consider  the following three cases, which  
reproduce different types of the FS: 
(a) $t^{(2)}/t^{(1)}=-1/6$, $t^{(3)}/t^{(1)}=0$, 
(b) $t^{(2)}/t^{(1)}=0$, $t^{(3)}/t^{(1)}=0$, 
and (c) $t^{(2)}/t^{(1)}=-1/6$, 
$t^{(3)}/t^{(1)}=1/5$.  
The case (a) is just what we have considered, and will be used 
as a reference below. 

Figure \ref{FSs} shows the FSs 
for each case at high temperature ($T=0.2 J$) in the {\it isotropic} 
$t$-$J$ model. 
The $\D$-dependence of $T_{\rm q1d}$ obtained 
under the constraint $\Delta_{\tau}=0$ 
is shown in \fig{Tq1ds}. 
The value of $\D_{\rm q1d}$  depends strongly on the band parameters, and 
is about (a) 0.13, (b) 0.075, and (c) 0.04, respectively. 
The q-1d state is most favored for case (a) 
because, as shown in \fig{FSs}, the FS  is 
located near $(\pi,\,0)$ and $(0,\,\pi)$  compared to 
the other cases, especially at low $\D$. 
Although the realistic $\D$ for high-$T_{\rm c}$ cuprates 
may be at most $0.30$, we note for case (c) that  
the q-1d instability occurs again at $\D \approx 0.46$-$0.48$ with 
$T_{\rm q1d} \slt 0.008 J$.  
This is because the FS passes near the points 
$(\pi,\,0)$ and $(0,\,\pi)$ around $\D \sim 0.45$. 

On the other hand, when we remove the constraint 
$\Delta_{\tau}=0$ in the {\it isotropic} 
$t$-$J$ model, the $d$-RVB state completely overcomes 
the q-1d state. This feature is common to the three cases. 

In the {\it anisotropic} $t$-$J$ model with $\theta = 5^{\circ}$, 
we observe that the anisotropy 
$\frac{\chi_{x}-\chi_{y}} 
{\chi_{x}+\chi_{y}}$ forms the cusp structure 
as a function of $T$ 
in a region below $\D \sim$   
(a) $0.20$, (b) $0.15$, and (c) $0.10$, respectively. 
This band parameter dependence 
reflects the different value of $\D_{\rm q1d}$ for each case. 
For case (c), however, the cusp structure reappears 
above $\D \sim 0.35$. In addition, the value of 
$\frac{\chi_{x}-\chi_{y}} 
{\chi_{x}+\chi_{y}}$  
at $T\approx 0$ increases with $\D$ above $\D \approx 0.15$-$0.20$ while 
it decreases with $\D$ for the other cases as shown in \fig{Xq1ds}(a). 
These different behaviors for case (c) 
can be understood as due to the proximity of the FS to the points  
$(\pi,\,0)$ and $(0,\, \pi)$ at the higher $\D$.

\section{Discussion}
\subsection{Comparison with experiments}
Now we discuss a relevance of the present  q-1d state to 
 high-$T_{\rm c}$ cuprates. 
The constraint $\Delta_{\tau}=0$ should be removed in the discussion. 
The results in the preceding section indicate  
two important factors: 
(i) a spatial anisotropy in $t^{(1)}$ and $J$, 
and (ii)  the values of $t^{(1)}$, $t^{(2)}$ and $t^{(3)}$. 
The former has effectively exposed the q-1d instability which was hidden 
behind the $d$-RVB state as shown in \fig{Xq1ds}; 
the extent of the \lq stability region' of the q-1d state 
can  be roughly measured by the value of $\D_{\rm q1d}$ 
as discussed in \S3.2 and \S3.3. 
This value of $\D_{\rm q1d}$ strongly depends on the latter factor.

\subsubsection{La$_{2-x}$Sr$_{x}$CuO$_4$}
For LSCO, we take band parameters, $t^{(1)}/J=4, t^{(2)}/t^{(1)}=-1/6$ and 
$t^{(3)}/t^{(1)}=0$. This choice reproduces the observed FS at
$\D=0.30$\cite{ino} in the {\it isotropic} $t$-$J$ model. 

We first discuss La$_{1.6-x}$Nd$_{0.4}$Sr$_{x}$CuO$_4$, assuming the same
band parameters as those of LSCO.
The crystal structure is LTT or $Pccn$ (an intermediate structure between
 LTO and LTT) at temperatures below $T_{\rm d2}$
in a range $0\slt \D \slt 0.30$\cite{crawford,buchner},
and the static spatial anisotropy is present in $t^{(1)}$ and $J$.
 We thus expect the realization of the static q-1d state
below $T_{\rm d2}$ or its coexistence with the $d$-RVB.
 Even above $T_{\rm d2}$, the dynamical q-1d fluctuations is expected
as discussed below.

On the other hand, for LSCO the crystal structure 
is LTO and hence allows no 
static spatial anisotropy in $t^{(1)}$ and $J$. 
 The use of the results for the {\it anisotropic} $t$-$J$ model
obtained in the preceding section is thus not justified.
 However, noting the existence of the Z-point soft phonon mode
associated with the structural phase transition from LTO to LTT
at low temperature in a range
$0 \leq \D \leq 0.18$\cite{thurston,chlee,kimura},
we expect a spatial anisotropy in $t^{(1)}$ and $J$
within a time scale $\omega_{\rm ph}^{-1}$ and a spatial scale of the 
correlation length  of the LTT fluctuation, 
where $\hbar \omega_{\rm ph} = 1$-$2$ meV is the energy
of the Z-point soft phonon mode (called as the \lq LTT-phonon' below).
 To estimate the value of $\theta$, we recall 
an experimental indication\cite{axe} 
that the LTT fluctuation around the LTO
structure occurs as a simple rotation of the CuO$_6$ tilting
direction in the plane, namely, from {\it e.g.} [110] to {\it e.g.}
 [100] (tetragonal notation), 
as successfully modeled by a classical XY model.
  This means that the magnitude of the (instantaneous) LTT
distortion can be as large as that of the (time-averaged)
LTO distortion.
 Since the tilting angle in the LTO structure 
is $\theta \approx 2$-$5^{\circ}$
for $\delta < 0.18$\cite{radaelli}, our choice of  $\theta = 5^{\circ}$ for
the  LTT distortion will be reasonable in magnitude.
 Taken these, we propose that in LSCO with the \lq LTT-phonon' 
the q-1d state (or its coexistence with the $d$-RVB) is realized as
dynamical fluctuations within time scales shorter than
$\omega_{\rm ph}^{-1}$.
 Since the CuO$_{6}$ tilting pattern of the \lq LTT-phonon' 
alternates between the $x$- and $y$-directions along the $c$-axis,
the q-1d state (or precisely, q-1d fluctuations) will also have the same
alternate structure (or alternate correlations) along the $c$-axis.

 Because of the dynamical nature of the q-1d state in the LTO structure, 
the experimental observation of the proposed q-1d state will depend on
probes.
 High-energy probes ($\omega  \sgt \omega_{\rm ph}$), such as ARPES
and inelastic neutron scattering, will observe an 
instantaneous q-1d state, while low-energy probes 
($\omega \ll \omega_{\rm ph}$), such as NMR
and $\mu$SR, will observe a time-averaged state, which is 2d-like in
each CuO$_{2}$ plane.
 We have interpreted the data from the former class (ARPES and neutron)
in terms of the present q-1d picture\cite{yamase}. 
 Among others, we can fit the observed FS segments\cite{ino}  
semiquantitatively with the q-1d FSs determined in the present
{\it anisotropic} $t$-$J$ model with $\theta=5^{\circ}$  
at low temperature $(T \ll J)$. 

We note a recent report\cite{sakita} that LSCO has 
the $Pccn$ structure at low temperature at   $\D = 0.115$.    
According to the scenario so far described, 
the q-1d state can become static even in LSCO. 
In the reverse way, we may argue that 
the present coupling between (spin) fermions and phonons via the
anisotropy in $t^{(1)}$ and $J$ is the origin of the $Pccn$ structure
when  the q-1d fluctuations are frozen in the LTO structure.


\subsubsection{YBa$_{2}$Cu$_{3}$O$_{6+y}$}
Following the previous report\cite{tanamoto}, we take  
$t^{(1)}/J=4, t^{(2)}/t^{(1)}=-1/6$ and 
$t^{(3)}/t^{(1)}=1/5$. 
For $y \sgt 0.4$,  CuO chains order along the $b$-axis accompanying the 
orthorhombicity $(b-a)/(b+a) \slt 1$\% in the in-plane lattice
constants $a$ and $b$\cite{jorgensen}. 
(The crystal structure is tetragonal for $y \slt 0.4$.) 
 A weak coupling to the CuO chain 
band will cause the spatial anisotropy, 
$t_{y}^{(1)}/t_{x}^{(1)} >1$, which 
will be, however, reduced by the orthorhombicity whose effect is 
estimated as\cite{harrison}  
$t_{y}^{(1)}/t_{x}^{(1)} \propto (\frac{a}{b})^{3.5} <1 $. 
The resulting anisotropy may be comparable to or less than 
that in LSCO. In addition, with the present choice of band parameters 
the degree of the 
intrinsic q-1d instability is very small compared 
to the case of LSCO (\fig{Tq1ds}).  
Figure \ref{Yq1d} indeed shows that 
the FSs for $\D=0.05$ and $0.30$  remain almost 2d at $T=0.01 J$ in the 
{\it anisotropic} $t$-$J$ model with $\theta =5^{\circ}$. 
(Such a  parametrization in terms of 
$\theta$ is, of course,  not appropriate for YBCO, where there is no 
\lq tilting'. Hence, the use of $\theta$ is just for 
convenience in a comparison with the case of LSCO.) 
Therefore YBCO system is not effective in realizing the q-1d state, 
and instead the 2d $d$-RVB state will be 
realized at low temperature. 
This picture is consistent with the ARPES data\cite{schabel} in that 
the observed FS at $T \sim 20$K is 2d hole-like 
centered at $(\pi,\,\pi)$.

\subsection{Possible charge inhomogeneity}
We have assumed that the charge (boson) distribution 
is homogeneous. If we relax this restriction, 
it is possible that the charge distribution 
becomes inhomogeneous and especially takes a q-1d structure 
in the state with the q-1d FS. 
In this connection, 
the \lq charge stripe' picture\cite{tranquada1,tranquada2} 
will be interesting. These aspects, including the possible 
competition with the Bose condensation or superconductivity, 
are left to future studies.

\subsection{Nearest neighbor Coulomb interaction}
As seen in \S 3.2, a small perturbation to the 
original {\it isotropic} $t$-$J$ model has exposed 
its intrinsic q-1d instability. From the same viewpoint, 
the role of the n.n. Coulomb interaction, $V$, will be  interesting. 
Our preliminary calculation in the {\it isotropic} $t$-$J$ model with 
$t^{(1)}/J =4$, $t^{(2)}/t^{(1)}=-1/6$ and $t^{(3)}/t^{(1)}=0$ shows that 
a reasonable value of $V$ stabilizes 
the coexistence of the q-1d state with  the $d$-RVB 
below $\D \sim 0.10$\cite{yamase4}. 
Therefore, in realizing  the q-1d state, effects of $V$ are  cooperative 
with those of the small spatial anisotropy in $t^{(1)}$ and $J$, and 
the former 
tends to freeze the q-1d fluctuation due to the \lq LTT-phonon'.

\section{Summary}
We have found within the slave-boson mean field approximation that  
the 2d $t$-$J$ model has an intrinsic instability toward forming a q-1d FS. 
This q-1d instability is driven mainly by fermions on the 
FS near $(\pi,\,0)$ and $(0,\, \pi)$, and thus 
competes with the $d$-RVB. 
For a realistic parameter choice, 
the $d$-RVB state completely overcomes the q-1d state. 
However, we have shown that 
a small spatial anisotropy in $t^{(1)}$ and $J$ exposes the  
q-1d instability which has been  hidden behind  the $d$-RVB state, 
and brings about the coexistence with the $d$-RVB. 
We have argued that this coexistence can
be realized in LSCO systems.  

\vspace{2cm}

\textit{Acknowledgements.} 
We thank Professor H. Fukuyama for his continual encouragement. 
H. Y. also thanks Professor T. Fujita for informing him of 
ref. 22. 
This work is supported by a Grant-in-Aid for Scientific Research from 
Monbusho. 

\vspace{2cm}

\bfi[h]
\bc
\includegraphics[width=5cm]{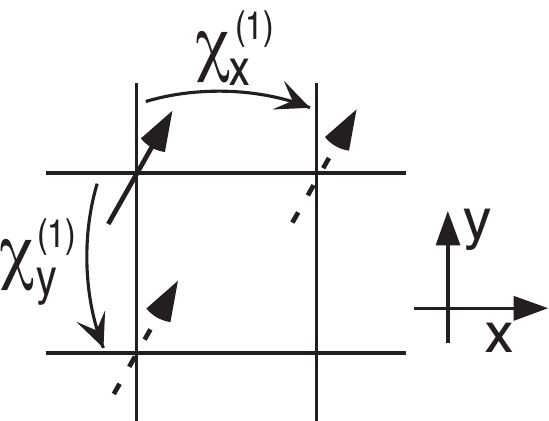}
     \caption{Fermion hopping amplitudes, $\chi_{x}^{(1)}$ and  
$\chi_{y}^{(1)}$, central quantities in this paper. They are 
abbreviated to $\chi_{x}$ and $\chi_{y}$, respectively.}
    \label{Xij}
    \ec
\efi

 \bfi
 \bc
 \includegraphics[width=8cm]{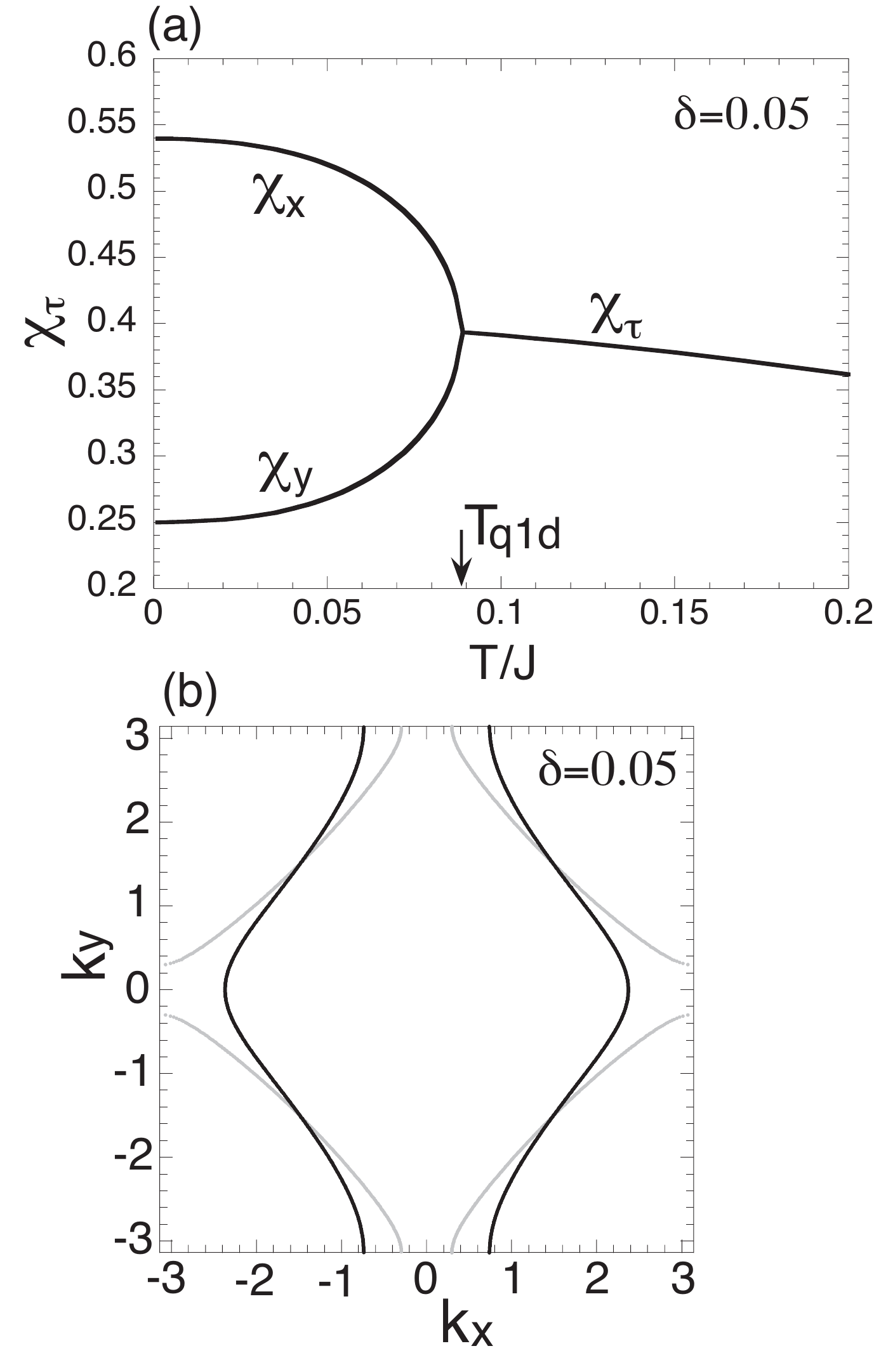}
   \caption{(a) $T$-dependence of 
$\chi_{\tau}$ in the {\it isotropic} 
$t$-$J$ model with the constraint 
$\Delta_{\tau}=0$.  The four-fold 
symmetry is  broken spontaneously below $T_{\rm q1d}$, that is  
$\chi_{x} \neq \chi_{y}$. 
(b) Fermi surface for $T > T_{\rm q1d}$ (gray line) and that for 
$T< T_{\rm q1d}$ (solid line).} 
    \label{Xq1d}
    \ec
\efi

\bfi
\bc
 \includegraphics[width=8cm]{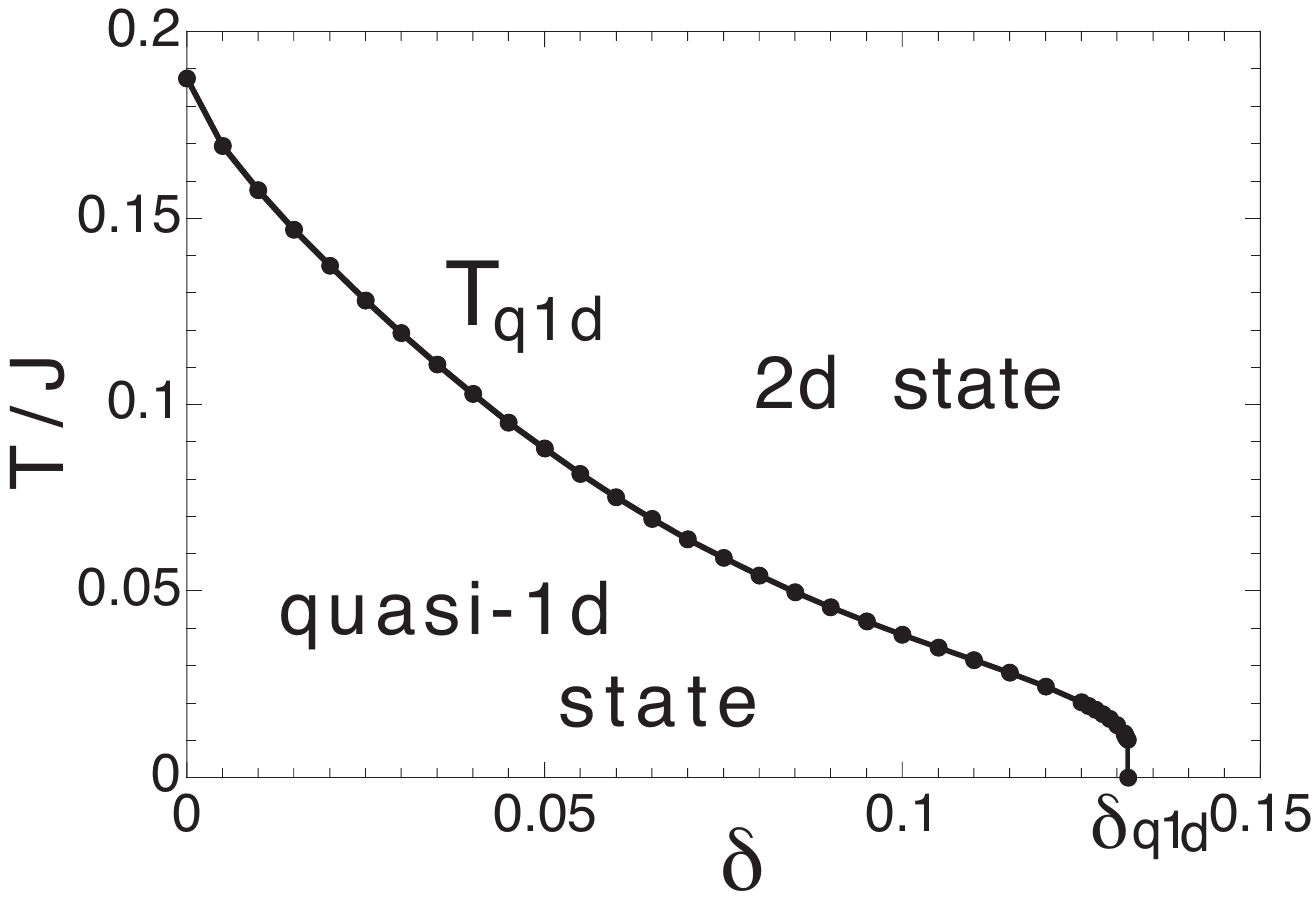}
   \caption{$\D$-dependence of $T_{\rm q1d}$ 
in the {\it isotropic} $t$-$J$ model 
with the constraint $\Delta_{\tau}=0$. \qquad \qquad \qquad 
\qquad \qquad \qquad \qquad \qquad }  
    \label{Tq1d}
    \ec
\efi

\bfi
\bc
 \includegraphics[width=8cm]{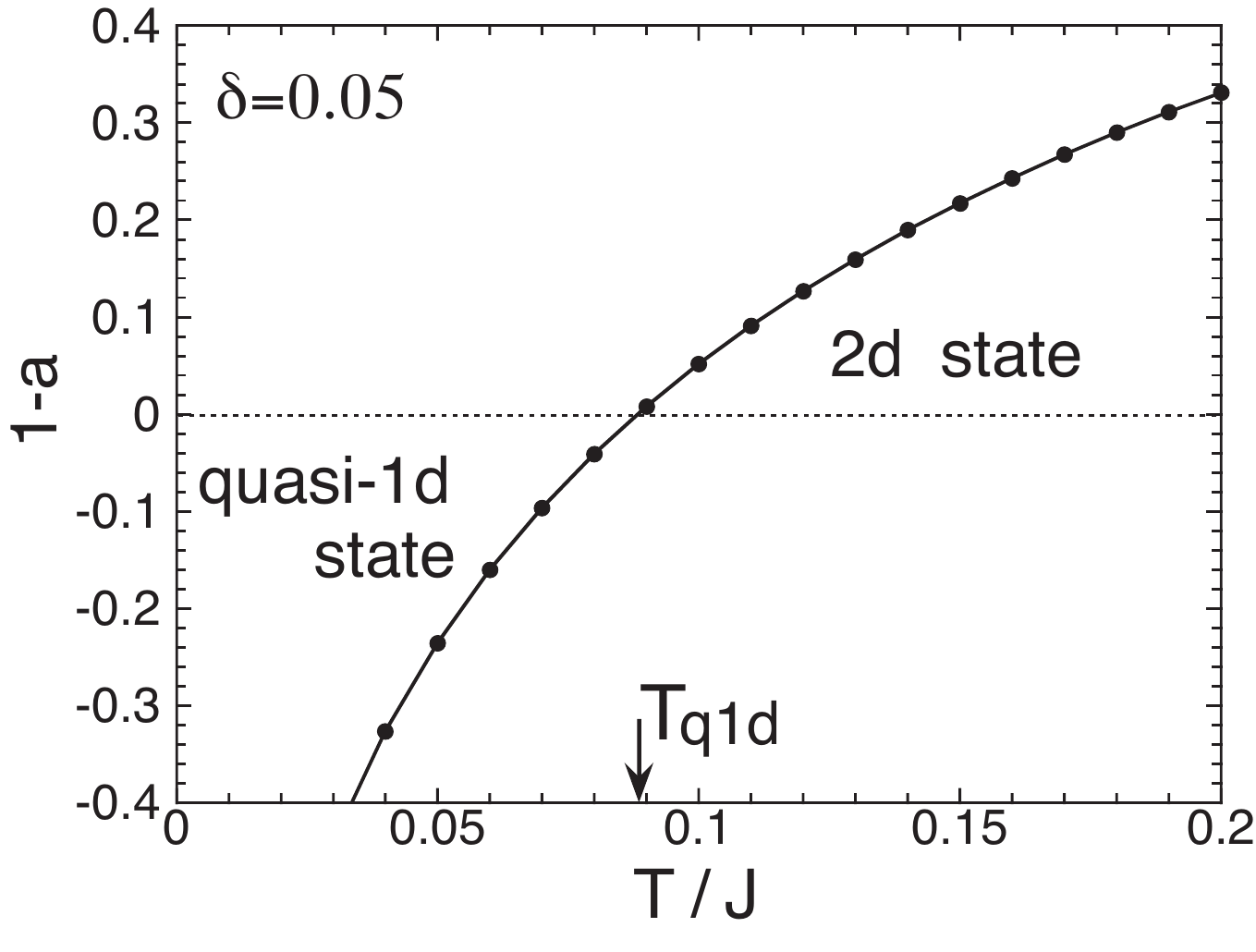}
   \caption{GL coefficient, $1-a$, for several $T$ at 
$\D=0.05$ under the constraint 
$\Delta_{\tau}=0$. It becomes negative below  $T_{\rm q1d}\approx 0.09J$, 
signaling an instability toward the  quasi-1d state.}
    \label{1-a}
    \ec
\efi

\bfi 
\bc
 \includegraphics[width=8cm]{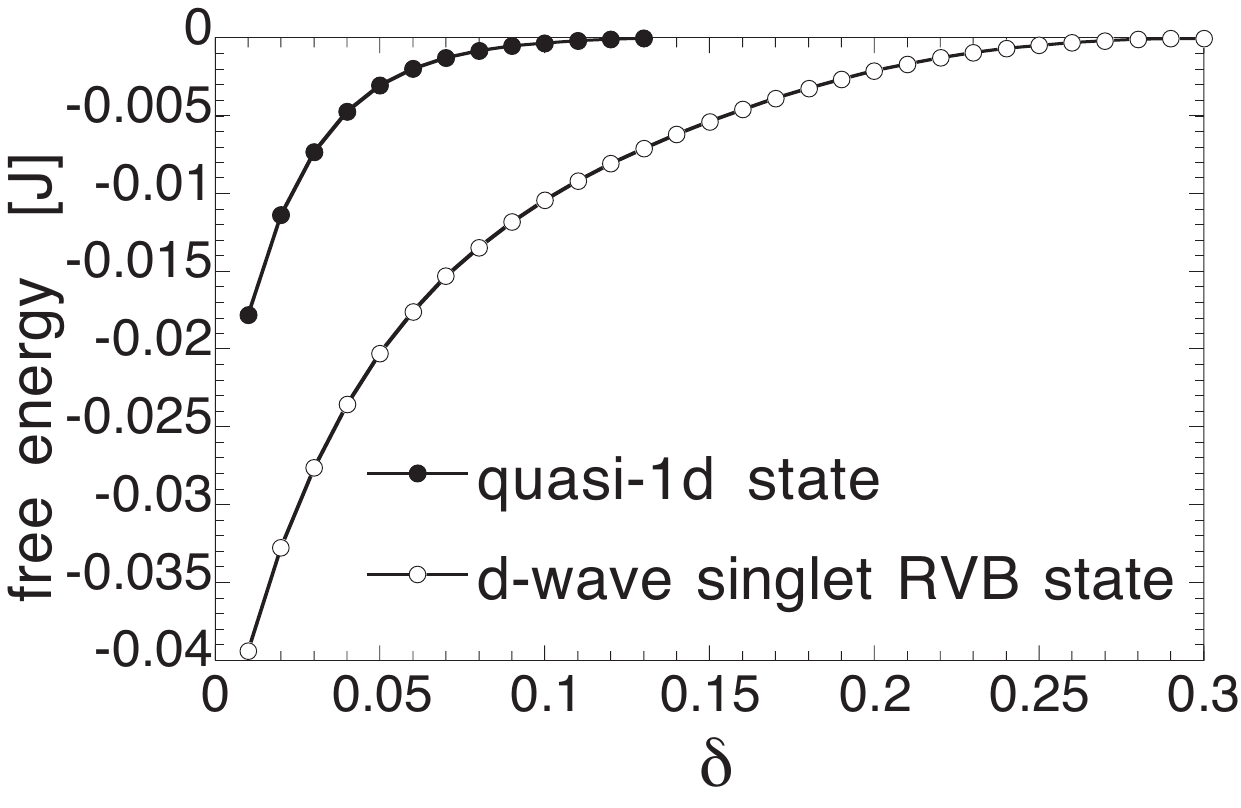}
 \caption{Free energy at $T=0.01 J$ of the quasi-1d state and of 
the $d$-wave singlet RVB ($d$-RVB) state,   
relative to that of  the isotropic 2d state without the $d$-RVB.} 
 \label{energy gain}
\ec
\efi

\bfi 
\bc
 \includegraphics[width=9cm]{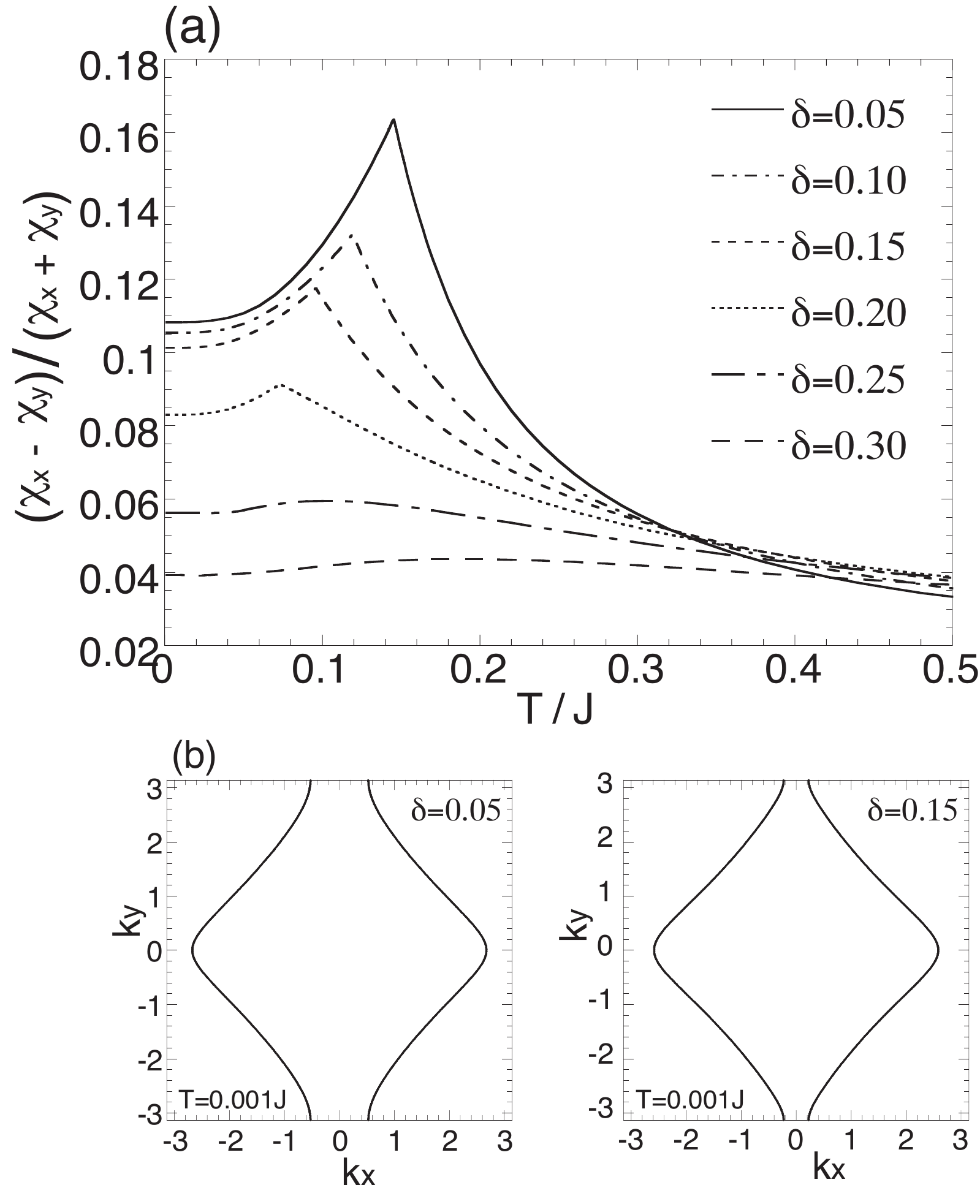}
 \caption{(a) $T$-dependence of the degree of the anisotropy, 
$\frac{\chi_{x}-\chi_{y}}
{\chi_{x}+\chi_{y}}$,  
for several choices of $\D$ in the {\it anisotropic} $t$-$J$ model with 
$\theta = 5^{\circ}$.  
(b) Quasi-1d Fermi surfaces in a state coexistent with the 
$d$-RVB at $T=0.001 J$ for $\D =0.05$ and 
$0.15$. The Fermi surface is defined by 
$\xi_{\vk}=0$, although the fermion dispersion is given by  
$E_{\vk}=\sqrt{\xi_{\vk}^{2}+ |\Delta_{\vk}|^{2}}$ in the 
coexistent state.}
 \label{Xq1ds}
 \ec
\efi

\bfi 
\bc
 \includegraphics[width=8cm]{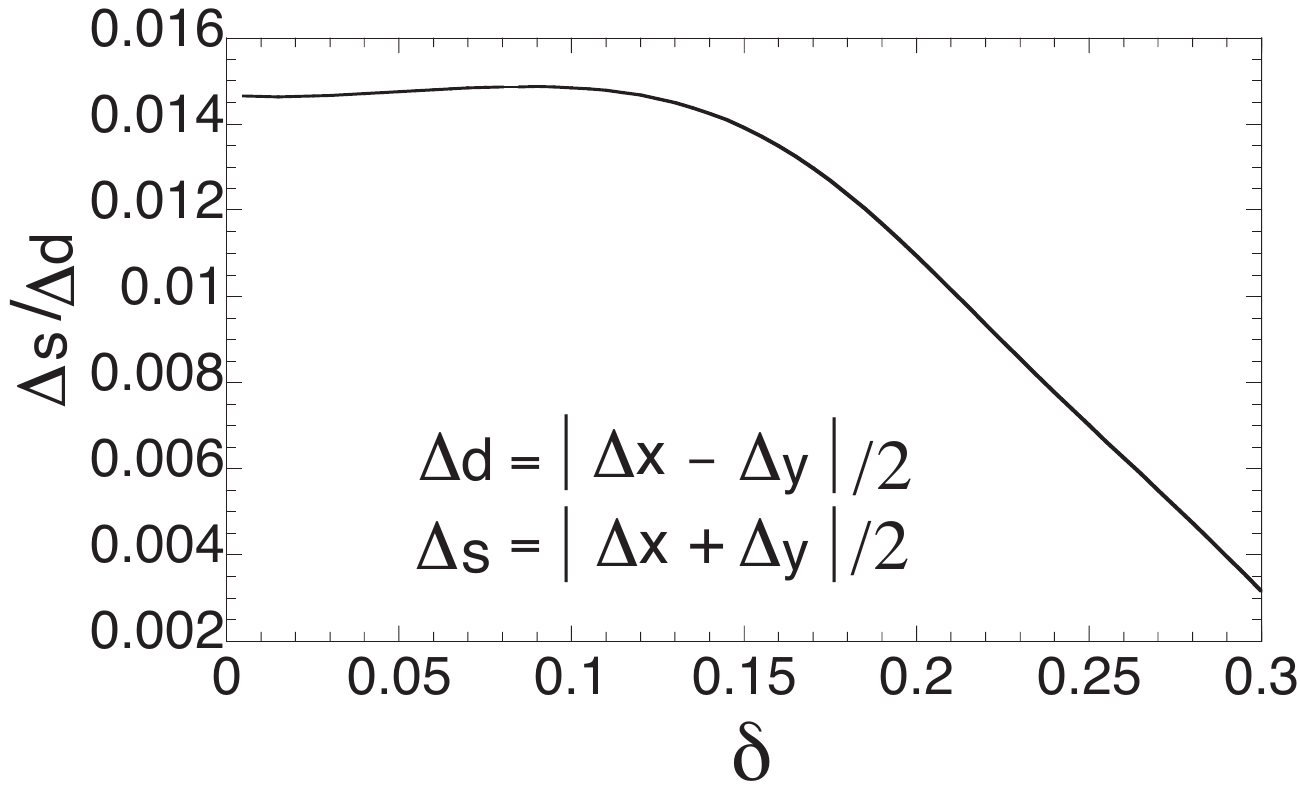}
 \caption{Relative magnitude of the extended $s$-wave component, 
$\Delta_{s}/\Delta_{d}$, as a function of $\D$ at $T=0.001 J$. 
The $\theta$ is set to $5^{\circ}$ in the {\it anisotropic} $t$-$J$ model.}
 \label{s-wave}
\ec
\efi

\bfi 
\bc
 \includegraphics[width=15cm]{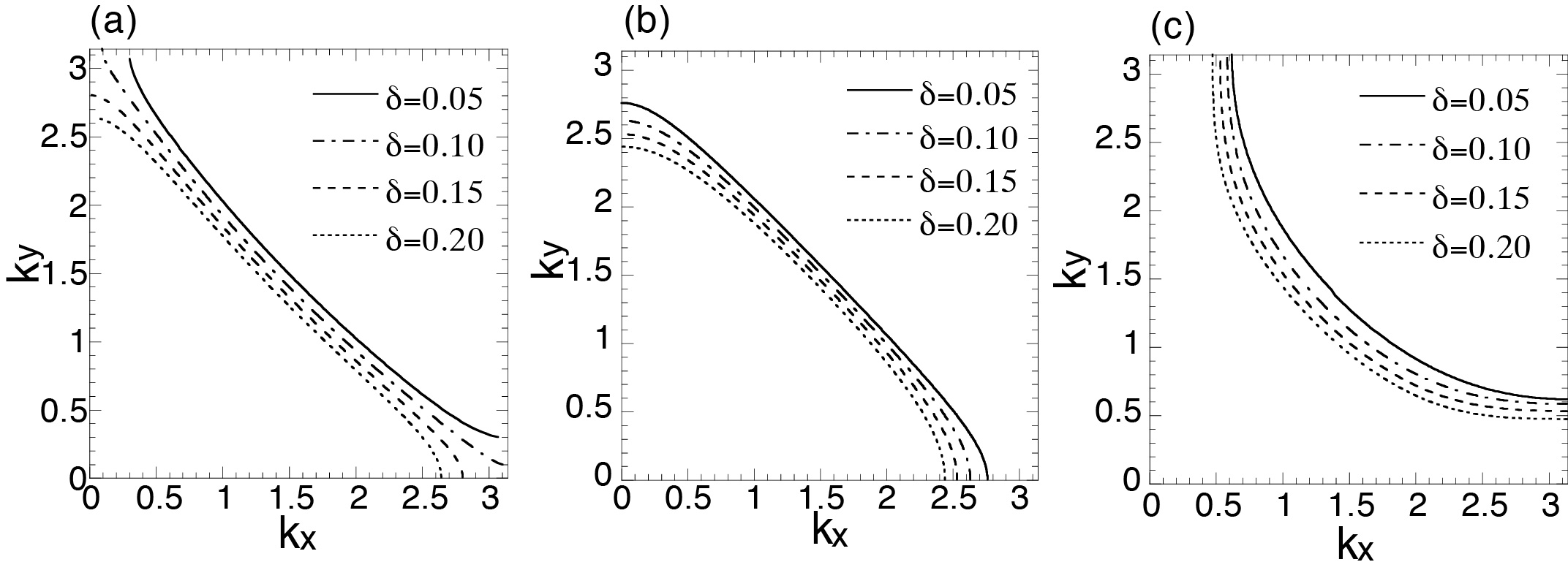}
 \caption{Fermi surfaces at $T=0.2 J$ 
in the {\it isotropic} $t$-$J$ model:  
(a) $t^{(2)}/t^{(1)}=-1/6$, $t^{(3)}/t^{(1)}=0$, 
(b) $t^{(2)}/t^{(1)}=0$, $t^{(3)}/t^{(1)}=0$, 
and (c) $t^{(2)}/t^{(1)}=-1/6$, $t^{(3)}/t^{(1)}=1/5$. 
They have a four-fold symmetry.} 
 \label{FSs}
 \ec
\efi

\bfi 
\bc
 \includegraphics[width=8cm]{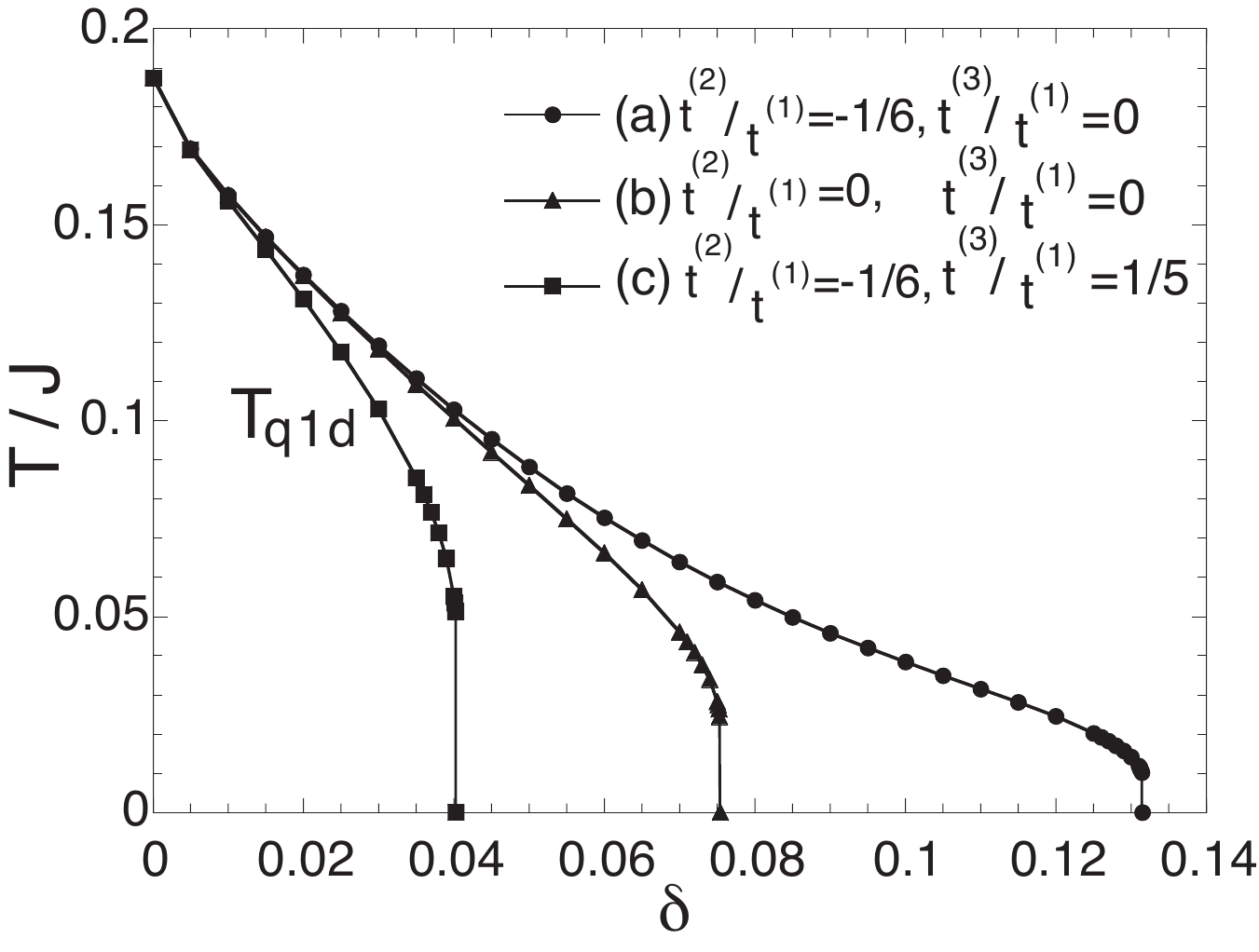}
 \caption{$T_{\rm q1d}$ as a function of $\D$ 
in the {\it isotropic} $t$-$J$ model under the 
constraint $ \Delta_{\tau} =0$: 
(a) $t^{(2)}/t^{(1)}=-1/6$, $t^{(3)}/t^{(1)}=0$, 
(b) $t^{(2)}/t^{(1)}=0$, $t^{(3)}/t^{(1)}=0$, 
and (c) $t^{(2)}/t^{(1)}=-1/6$, $t^{(3)}/t^{(1)}=1/5$.  
The values of $\D_{\rm q1d}$ are about (a) 0.13, (b) 0.075, and (c) 0.04, 
 respectively.}
 \label{Tq1ds}
 \ec
\efi

\bfi 
\bc
 \includegraphics[width=8cm]{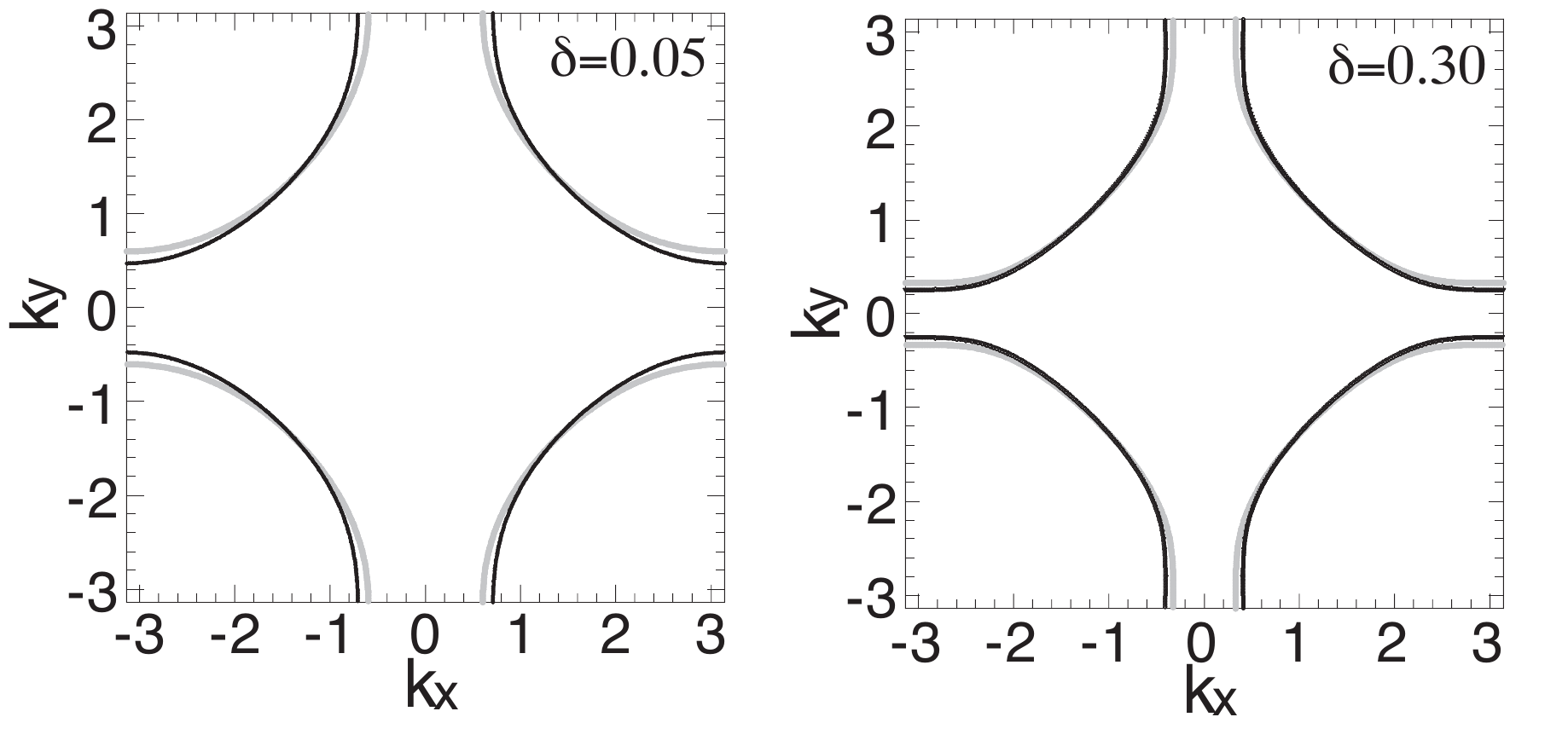}
 \caption{Fermi surfaces  for $\D=0.05$ and $0.30$ 
at $T=0.01J$ in  the {\it anisotropic} $t$-$J$ model with 
$\theta=5^{\circ}$ (solid line) and $0^{\circ}$ (gray line). 
The band parameter is taken as $t^{(1)}/J =4$, $t^{(2)}/t^{(1)}=-1/6$ and
$t^{(3)}/t^{(1)}=1/5$, appropriate to YBCO. 
The Fermi surface is defined by 
$\xi_{\vk}=0$, although the fermion dispersion is given by  
$E_{\vk}=\sqrt{\xi_{\vk}^{2}+ |\Delta_{\vk}|^{2}}$.} 
 \label{Yq1d}
\ec
\efi


\begin{thebibliography}{99}     


\bibitem{tranquada1} J. M. Tranquada, B. J. Sternlieb, J. D. Axe, Y. Nakamura 
                    and S. Uchida: Nature {\bf 375} (1995) 561.    
\bibitem{tranquada2} J. M. Tranquada, J. D. Axe, N. Ichikawa, 
                    Y. Nakamura, S. Uchida and B. Nachumi: \PR{54}  
                     (1996) 7489.
\bibitem{tranquada3} J. M. Tranquada, J. D. Axe, N. Ichikawa, 
                     A. R. Moodenbaugh, Y. Nakamura and S. Uchida: \PRL{78}  
                     (1997) 338.


\bibitem{white1} S. R. White and D. J. Scalapino: \PRL{80} (1998) 1272. 
\bibitem{white2} S. R. White and D. J. Scalapino: \PRL{81} (1998) 3227. 
\bibitem{white3} S. R. White and D. J. Scalapino: cond-mat/9907243.
\bibitem{hellberg1} C. S. Hellberg and E. Manousakis: \PRL{83} (1999)
 132. 
\bibitem{hellberg2} C. S. Hellberg and E. Manousakis: cond-mat/9910142.

\bibitem{yamase} H. Yamase and H. Kohno: \JPSJ{69} (2000)
 332; H. Yamase, H. Kohno and H. Fukuyama: Physica B {\bf 284-288}
	(2000) 1375. 

\bibitem{ino} A. Ino, C. Kim, T. Mizokawa, Z.-X. Shen, A. Fujimori, 
	M. Takaba, K. Tamasaku, H. Eisaki and S. Uchida: \JPSJ{68} (1999) 
	1496.  
\bibitem{yamada} K. Yamada, C. H. Lee, K. Kurahashi, J. Wada, S. Wakimoto,                  S. Ueki, H. Kimura, Y. Endoh, S. Hosoya, G. Shirane, 
                 R. J. Birgeneau, M. Greven, M. A. Kastner and Y. J. Kim:
                 \PR{57} (1998) 6165. 
\bibitem{yamase3} H. Yamase, H. Kohno and H. Fukuyama: 
to appear in Physica C. 

\bibitem{tanamoto} T. Tanamoto, H. Kohno and H. Fukuyama: 
                   \JPSJ{62} (1993) 717. 
\bibitem{xD} We consider $\D$ to be equal to the Sr$^{2+}$ content, $x$.  


\bibitem{bruce} B. Normand, H. Kohno and H. Fukuyama: \PR{53} (1996) 
                856. 
\bibitem{yamase2} H. Yamase, H. Kohno, H. Fukuyama and M. Ogata: 
                 \JPSJ{68} (1999) 1082. 
\bibitem{radaelli} P. G. Radaelli,  
D. G. Hinks, A. W. Mitchell,
 B. A. Hunter, J. L. Wagner, B. Dabrowski, K. G. Vandervoort,
 H. K. Viswanathan and J. D. Jorgensen: \PR{49} (1994) 4163. 
\bibitem{thurston} T. R. Thurston,  
R. J. Birgeneau, D. R. Gabbe,
 H. P. Jenssen, M. A. Kastner, P. J. Picone, N. W. Preyer, J. D. Axe,
 P. B\"{o}ni, G. Shirane, M. Sato, K. Fukuda and S. Shamoto: \PR{39}
 (1989) 4327. 
\bibitem{chlee} C. H. Lee, K. Yamada, M. Arai, S. Wakimoto, S. Hosoya and 
	Y. Endoh: Physica C {\bf 257} (1996) 264. 
\bibitem{kimura} H. Kimura, K. Hirota, C. H. Lee, K. Yamada and
 G. Shirane: cond-mat/9908217. 

\bibitem{axe} J. D. Axe, A. H. Moudden, D. Hohlwein, D. E. Cox, 
K. M. Mohanty, A. R. Moodenbaugh and Youwen Xu: 
\PRL{62} (1989) 2751. 


\bibitem{sakita} S. Sakita, F. Nakamura, T. Suzuki and T. Fujita: 
\JPSJ{68} (1999) 2755. 

\bibitem{crawford} M. K. Crawford, R. L. Harlow, E. M. McCarron,
 W. E. Farneth, J. D. Axe, H. Chou and Q. Huang: \PR{44} (1991) 7749. 
\bibitem{buchner} B. B\"{u}chner, M. Breuer, A. Freimuth and
 A. P. Kampf: \PRL{73} (1994) 1841. 



\bibitem{jorgensen} J. D. Jorgensen, B. W. Veal, A. P. Paulikas,
 L. J. Nowicki, G. W. Crabtree, H. Claus and W. K. Kwok: 
\PR{41} (1990) 1863. 

\bibitem{harrison} W. A. Harisson: {\it Electronic Structure and 
            the Properties of Solids} (Freeman, New York, 1980). 

\bibitem{schabel} M. C. Schabel, C.-H. Park, A. Matsuura, Z.-X. Shen, 
D. A. Bonn, Ruixing Liang and W. N. Hardy: \PR{57} (1998) 6107. 
\bibitem{yamase4} H. Yamase and H. Kohno: in preparation. 



\end{thebibliography}
\end{document}